\documentclass[aps,prb,floatfix,superscriptaddress,noshowpacs,twocolumn]{revtex4-2}
\usepackage{graphics,graphicx,epsfig,amsmath,amssymb,epstopdf,wasysym,comment}
\usepackage{hyperref}
\usepackage[dvipsnames]{xcolor}
\usepackage{braket,amsfonts}
\usepackage{dsfont}

\allowdisplaybreaks

\newcommand{\Pc}{\Psi^\dagger}

\newcommand{\be}{\begin{equation}}
\newcommand{\ee}{\end{equation}}
\newcommand{\bga}{\begin{gather}}
\newcommand{\ega}{\end{gather}}
\newcommand{\bea}{\begin{eqnarray}}
\newcommand{\eea}{\end{eqnarray}}
\newcommand{\dagga}{{\phantom{\dagger}}}

\newcommand{\bQ}{\mathbf{Q}}

\newcommand{\bK}{\mathbf{K}}
\newcommand{\bG}{\mathbf{G}}

\newcommand{\bq}{\mathbf{q}}

\newcommand{\bk}{\mathbf{k}}

\newcommand{\up}{\uparrow}
\newcommand{\down}{\downarrow}

\newcommand{\eqn}[1]{(\ref{#1})}

\newcommand{\ep}{{\epsilon}}

\newcommand{\bw}{\begin{widetext}}
\newcommand{\ew}{\end{widetext}}
\newenvironment{eqs}%
{\begin{equation} \begin{aligned}}%
{\end{aligned} \end{equation} }
\newcommand{\beal}{\begin{eqs}}
\newcommand{\eal}{\end{eqs}}
\newcommand{\bd}[1]{{\boldsymbol{#1}}}
\newcommand{\esp}[1]{\text{e}^{#1}}

\begin{document}
\title{Local Kekul\'e distortion turns twisted bilayer graphene into topological %valence-bond 
Mott insulators and superconductors}

\author{Andrea Blason}
\affiliation{International School for Advanced Studies (SISSA), Via Bonomea 265, I-34136 Trieste, Italy} 
\author{Michele Fabrizio}
\affiliation{International School for Advanced Studies (SISSA), Via Bonomea 265, I-34136 Trieste, Italy} 

\begin{abstract}
Magic-angle twisted bilayer graphene displays at different fillings of the four flat bands lying around the charge neutrality point a wealth of notable phases that include magnetic Chern insulators, whose magnetization is mostly of orbital nature, and contiguous superconducting domes. Such rich phase diagram is here explained through the positive interplay of Coulomb repulsion and the electron coupling to a twofold optical mode that corresponds to Kekul\'e distortions localized into the small AA stacked regions of the moir\'e supercells.  A static distortion stabilizes, at any integer filling of the flat bands, valence-bond insulators that carry finite Chern number away from charge neutrality. Similarly, a dynamic distortion that resonates between the two lattice vibrations leads to resonating-valence-bond topological insulators with built-in chiral d-wave pairs that have finite Chern number equal to the angular momentum, and thus are prone to turn superconducting upon doping away from integer filling. 
\end{abstract}

\maketitle 

\section*{Introduction}
Atomic relaxation in magic angle twisted bilayer graphene (TBLG) is responsible~\cite{Koshino-PRB2017,Mattia-PRB2018,Procolo-PRB2019,Guinea-relax-PRB2019,Kaxiras-PRR2019,Procolo-PRR2020}
of the gap opening between the four flat bands and all other upper and lower ones. Moreover, 
scattering by acoustic phonons has been invoked~\cite{DasSarma-phonons-PRB2019,Vignale-phonons-PRB2019} to explain the anomalous linear in temperature resistivity of the normal metal phases~\cite{Young-linear-T-NatPhys2019}, 
and both acoustic~\cite{Bernevig-phonons-PRL2019,DasSarma-phonons-PRB2019,DasSarma-AnnPhys2020,Lewandowski-PRB2021,Guinea-Cea-phonons-2021} and optical~\cite{MacDonald-phonons-PRL2018} modes have been explored as possible mechanisms of the observed superconductivity~\cite{Herrero-SC-Nature2018,Codecido-SC-ScAd2019,Yankowitz-SC-Science2019,Efetov-SC-Nature2019,Saito-SC-NatPhy2020,Balents-SC-NatPhys2020}. 
In spite of all that, the role of lattice degrees of freedom in the insulating phases of TBLG at integer fillings $\nu = \pm n$, $n=0,\dots,3$, of the flat bands~\cite{Herrero-MIT-Nature2018,Efetov-SC-Nature2019,Yankowitz-SC-Science2019,Andrei-Nature2019,Codecido-SC-ScAd2019,Sharpe-Science2019,Yazdani-Nature2019,Rubio-Nature2019,Saito-SC-NatPhy2020,Balents-SC-NatPhys2020,Serlin-Science2020,Nuckolls2020,Herrero-Nature2020} has been mostly overlooked in favour of Coulomb interaction~\cite{Vishwanath-PRX2018,Vafek-PRL2019,Xie-PRL2020,Zalatel-PRX2020,Zhang-PRB2020,XiDai-PRB2021,III:PhysRevB.103.205413,IV:PhysRevB.103.205414,Liao-PRX2021,PhysRevLett.122.246402}, which struggles to explain superconductivity~\cite{p+d-NPJ2019,PhysRevB.103.205415,Guinea-Cea-phonons-2021,Vishwanath-ScAdv2021}, and, especially, the anomalous quantum Hall effect and sizeable orbital magnetic moment~\cite{Efetov-arXiv2021,Yankowitz-arXiv2022} recently measured at $\nu=\pm 2$. 
%Moreover, screening the Coulomb interaction reduces the stability of the insulating states whereas the superconducting phase is strengthen, see Ref.~\cite{science.abb8754}, hindering a different mechanism in the formation of the Cooper pairs.

%%
\begin{figure}[t!]
\centerline{\includegraphics[width=\columnwidth]{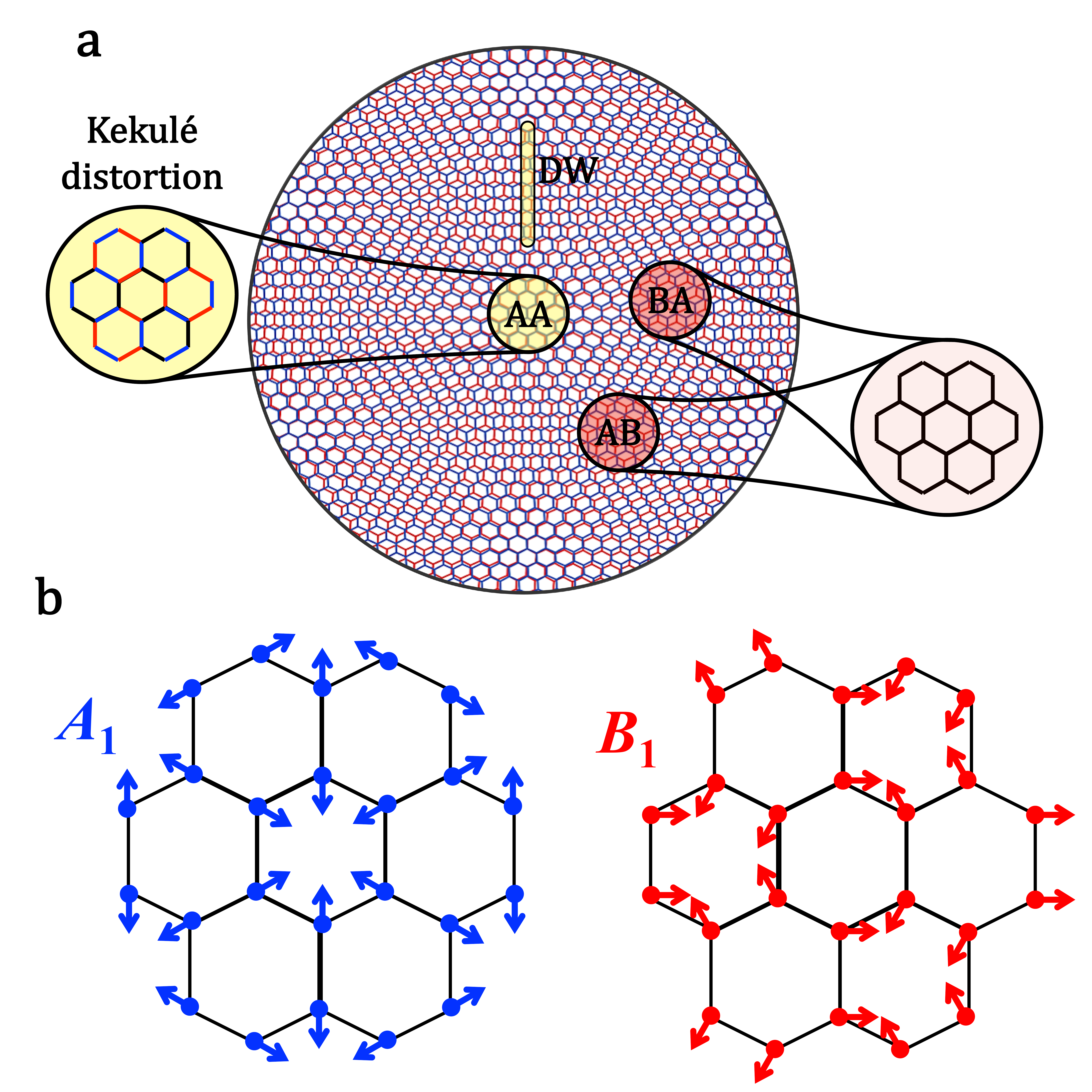}}
\caption{\textbf{Kekul\'e-like distortion that stabilises a non- magnetic insulator at charge neutrality}.
Panel \textbf{a}: Kekul\'e-like distortion that stabilises a non-magnetic insulator at charge neutrality. 
Such distortion is driven by moir\'e optical modes at the $\bd{\Gamma}$ point with $A_1$ and $B_1$ symmetry, which 
correspond to long-wavelength modulations of the single-layer graphene $A_1$ and $B_1$ modes shown in panel \textbf{b}. The moir\'e optical phonons, and thus the Kekul\'e distortion, mainly affect the AA stacked regions and the domain walls (DW), both shown in yellow in panel \textbf{a}, separating Bernal stacked AB and BA regions, shown in red, the latter remaining almost unaffected~\cite{Mattia-JT}. We emphasise that the lattice displacement occurs at the zone center of the reduced Brillouin zone and on the atomic scale of graphene, which distinguishes it from the Kekul\'e state discussed in  Ref.~\cite{Lee-PRB2018}.
}
\label{Kekule}
\end{figure}

The major importance of the electron-phonon coupling also emerged from  Ref.~\cite{Mattia-JT} that theoretically uncovered special, almost non dispersive, optical modes, later observed by nano-Raman spectroscopy~\cite{Gadelha-Nature2021},  which are so strongly coupled with the electrons that atomic displacements as small as 2m{\AA} are sufficient to open sizeable gaps in the flat bands at all integer $\nu$. Those phonons derive from the 1360~cm$^{-1}$ $A_1$ and $B_1$ TO modes of a single-layer graphene at the $\bK$ point, see Fig.~\ref{Kekule}\textbf{b}, superimposed with a long wavelength modulation driven by the van der Waals inter-layer interaction that makes these modes exist throughout the whole reduced Brillouin zone (RBZ) of the moir\`e superlattice and be localized into the AA stacked regions and the domain walls separating the AB and BA stacked ones. Remarkably, these special phonons have the same twofold accidental degeneracy of the flat bands along the $\bd{\Gamma}\to\bK\to\bd{M}$ high-symmetry path in the RBZ that reflects the emerging $U_v(1)$ valley symmetry~\cite{MacDonald,Mattia-JT}. Because of that, the $U_v(1)$ symmetric electron-phonon coupling effectively realises a Jahn-Teller model, which explains the efficacy of a static distortion on lifting the accidental degeneracy.

However, systematic theoretical studies of the lattice contribution to the phase diagram of TBLG are lacking. Indeed, Ref.~\cite{Mattia-JT} describes a realistic frozen-phonon tight-binding calculation that neglects Coulomb 
repulsion and can only access states with broken $U_v(1)$ and, eventually, broken spatial symmetries when the frozen-phonon is not at the $\bd{\Gamma}$ point,  but not the observed Chern insulators with spontaneously broken time-reversal symmetry.
Filling this gap is actually the scope of this work.
Specifically, upon integrating out phonons we obtain an effective electron-electron attraction that can be assumed instantaneous since the flat-band width is a lot smaller than the phonon frequency. We treat this interaction on an equal footing with Coulomb repulsion, investigating their mutual interplay and its effect on the phase diagram by means of Hartree-Fock and projected BCS-wavefunctions calculations.

Before discussing our findings, we believe worth placing them within the general context of correlation effects in graphene. We recall that interaction strength in graphene is sizeable but yet not enough to stabilise a correlated insulator at charge neutrality~\cite{Coulomb-graphene-PRL2011}. 
An isotropic strain above 8-10\% that expands all C-C bonds has been shown to stabilise~\cite{Sandro-PRL2018} both an antiferromagnetic insulator and a Kekul\'e valence-bond (KVB) one, with the latter lower in energy than the former. 
The Kekul\'e distortion involves just the above mentioned $A_1$ and $B_1$ modes of graphene, whose positive interplay with Coulomb repulsion thus favours the KVB insulator instead of the antiferromagnetic one expected from Coulomb repulsion alone. 
In light of the vanishingly small Fermi velocity at the Dirac cones in magic-angle TBLG, whose Bloch waves are primarily localised into the AA stacked regions just like the $A_1$ and $B_1$ moir\'e phonons, it is not unlikely that also in this case those phonons cooperate with Coulomb repulsion to stabilise a KVB insulator, with the distortion discussed in 
Ref.~\cite{Mattia-JT} and shown schematically in Fig.~\ref{Kekule}\textbf{a}.
% all the more since atomic relaxation~\cite{Koshino-PRB2017,Mattia-PRB2018,Procolo-PRB2019,Guinea-relax-PRB2019,Kaxiras-PRR2019,Procolo-PRR2020} naturally yields a weak tensile strain of the AA stacked regions.

That is precisely what we find within Hartree-Fock approximation. 
We hereafter denote such space-selective Kekul\'e distortion at the zone-center of the moir\'e Brillouin zone a static Kekul\'e valence bond (S-KVB) distortion.
The corresponding S-KVB insulator at charge neutrality seeds the cascade of symmetry-breaking mean-field insulating states at all other integer fillings.
\\
Since the AA regions are quite far apart from each other, as testified by the tiny dispersion of the $A_1$ and $B_1$ modes in TBLG~\cite{Mattia-JT}, it is well possible that a resonating rather than static Kekul\'e valence bond (R-KVB) insulator is stabilised, in which each AA region is instantaneously distorted along $A_1$ or $B_1$ but dynamically the symmetry is restored. 
\\
Since the electron-phonon coupling realises a Jahn-Teller model, S-KVB and R-KVB correspond to static and dynamic Jahn-Teller effect, respectively. 
Such R-KVB state thus effectively realises a Jahn-Teller-Mott insulator~\cite{Jahn-Teller-Mott}. The close analogy with Anderson's resonating valence bond scenario~\cite{PWA-RVB} for cuprates also suggests that the Jahn-Teller-Mott insulator is prone to become a superconductor upon doping; a 
phonon mediated superconductivity not hindered by Coulomb repulsion~\cite{PWA-RVB,Capone-PRL2004,Fabrizio-RMP2009}. We will show that such superconductor is likely to have chiral or nematic $d$-wave symmetry, in accordance with the 
analysis of Ref.~\cite{MacDonald-phonons-PRL2018} where the same TO phonons of graphene have been considered as driving mechanism of superconductivity in TBLG.  
\begin{figure}[t]
\centerline{\includegraphics[width=0.75\columnwidth]{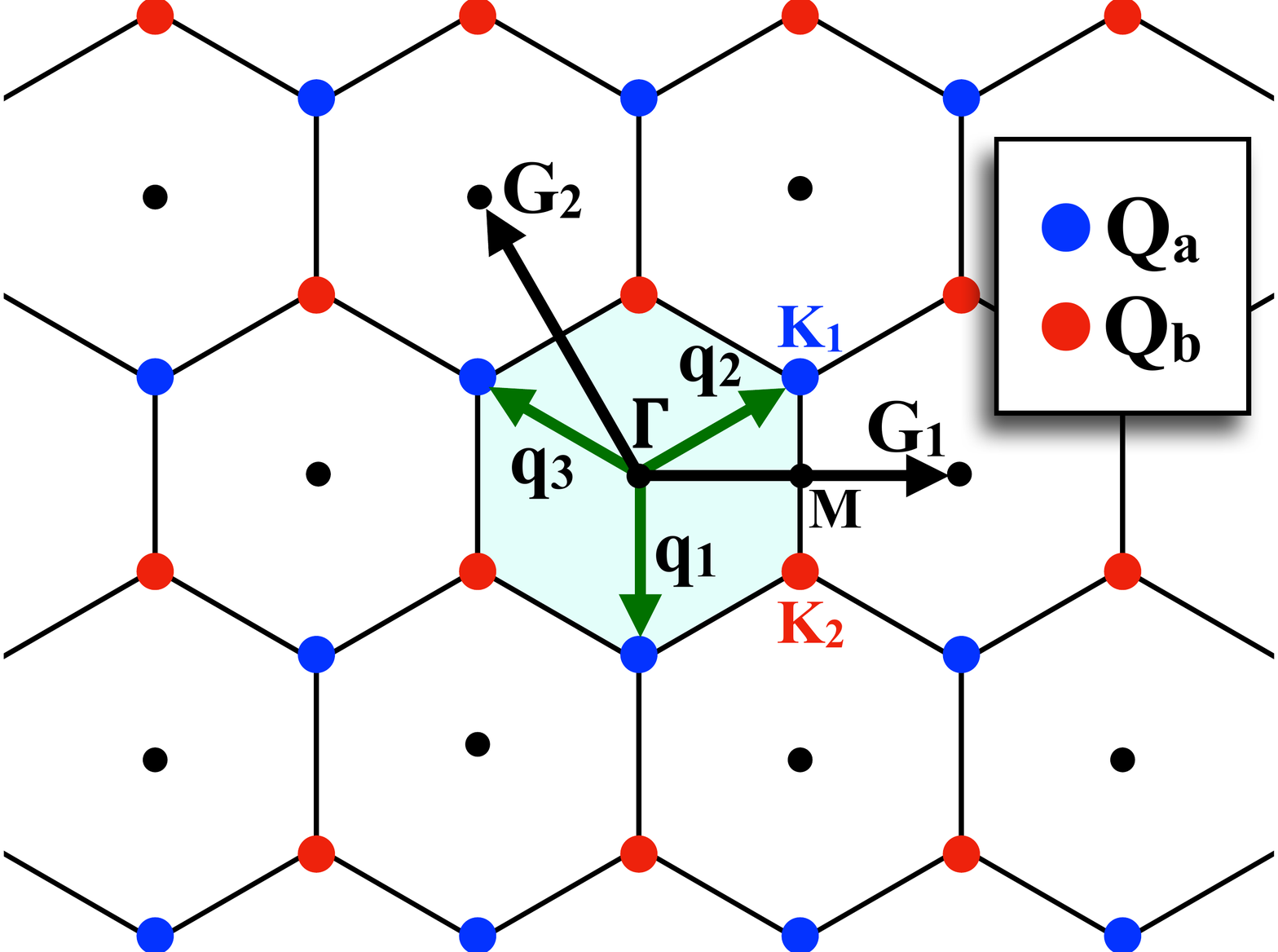}}
\caption{\textbf{Reciprocal lattice space}. The two sublattices  $\bQ_a$, blue dots, and $\bQ_b$, red dots, of the reduced Brillouin zones, in light cyan the first one, can be generated, e.g., through $\bQ_a=\bq_1+\bd{G}$ and 
$\bQ_b= -\bq_1+\bd{G}$, where $\bd{G}=n\bd{G}_1 + m\bd{G}_2$ is 
any reciprocal lattice vector with $\bd{G}_1$ and $\bd{G}_2$ the primitive ones. Also shown are the high symmetry points $\bd{\Gamma}$, $\bK_1$, $\bK_2$ and $\bd{M}$.}
\label{conventions}
\end{figure}

\section*{Model Hamiltonian and interaction}
We consider two AA stacked graphene layers, and rotate around the perpendicular axis layer 1 by $+\theta/2$ and layer 2 by $-\theta/2$, at magic angle $\theta=1.08^\circ$.  
We describe the band structure through the Bistritzer-MacDonald continuum model~\cite{MacDonald}, using the conventions of Ref.~\cite{Bernevig-PRL2019}. Specifically, we define four component spinors in momentum space $\Psi^\dagga_{\bk,\bQ_a,\sigma}$ and $\Psi^\dagga_{\bk,\bQ_b,\sigma}$, two components corresponding to sublattices $A$ and $B$ of a graphene layer, and the other two to the valley index $\eta=\pm1$, where $\sigma$ is the spin, $\bk$ runs within the first RBZ, while  
$\bQ_a$ and $\bQ_b$ identify the two sublattices in reciprocal space, see Fig.~\ref{conventions}. The operator $\Psi^\dagga_{\bk,\bQ_a,\sigma}$ is defined close to the Dirac point $\bK$ on layer 1 and $-\bK$ on layer 2 for $\eta=+1$ 
and $\eta=-1$, respectively, while in $\Psi^\dagga_{\bk,\bQ_b,\sigma}$ the two layers are interchanged. 
Moreover, the sublattice components of the spinors with $\eta=+1$ and $\eta=-1$ are inverted~\cite{Bernevig-PRL2019}. We implicitly assume that the longest reciprocal 
lattice vector kept in our calculation is still much smaller that the distance 
$2|\bK|$ between the two valleys, so that the chosen basis is not overcomplete. \\   
With those definitions, the non-interacting Hamiltonian can be written as 
\beal
H_0 &= \sum_{\bk\sigma}\,\sum_{\bQ,\bQ'}\,\Psi^\dagger_{\bk,\bQ,\sigma}\,
\hat{H}^{(0)}_{\bQ\bQ'}(\bk)\,\Psi^\dagga_{\bk,\bQ',\sigma}\,,
\eal
where $\bQ=\bQ_a\oplus\bQ_b$, and, in the zero-angle approximation~\cite{MacDonald}, 
\beal
&\hat{H}^{(0)}_{\bQ\bQ'}(\bk) = \delta_{\bQ,\bQ'}\,v_F\,\tau_3\,\big(\bk-\bQ\big)\cdot\bd{\sigma} +
\\&\qquad 
+\tau_0\, \sum_{i=1}^3\, \big(\delta_{\bQ-\bQ',q_i}+\delta_{\bQ'-\bQ,q_i}\big)\,
\hat{T}_i(u_0,u_1)\,.\label{H0}
\eal
Hereafter, the Pauli matrices $\tau_a$ and $\sigma_a$, $a=0,1,2,3$, act on the 
valley and sublattice indices, respectively, $\bq_i$, $i=1,2,3$, are defined in Fig.~\ref{conventions}, 
while
\beal
\hat{T}_1(u_0,u_1) &= u_0\,\sigma_0 + u_1\,\sigma_1\,,\\
\hat{T}_{j+1}(u_0,u_1) &= \esp{i\frac{2\pi}{3}\,\sigma_3}\;\hat{T}_j(u_0,u_1)\;
\esp{-i\frac{2\pi}{3}\,\sigma_3}\,, \qquad j=1,2\,. 
\label{T}
\eal
Setting as unit length the moir\'e primitive lattice vector for a twist angle of $1.08^{\circ}$, we fix
$v_F = 40 \, \text{meV}$, $u_0= 76.1 \, \text{meV}$ and $u_1= 103.1 \, \text{meV}$.\\
The charge density operators $\rho_\ell(\bq+\bd{G})$ 
of each layer $\ell=1,2$ 
are diagonal in sublattice and valley indices, and read, for $\bq\in\text{RBZ}$, 
\beal
&\rho_\ell(\bq+\bG) = \sum_{\bk\bQ\sigma}\,\Psi^\dagger_{\bk,\bQ,\sigma}\;
\hat{\rho}_\ell(\bQ)\,\sigma_0\;\Psi^\dagga_{\bk+\bq,\bQ-\bG,\sigma}\,,\\
&\hat{\rho}_\ell(\bQ) = \delta_{\bQ,\bQ_a}\,\frac{\tau_0-(-1)^\ell\,\tau_3}{2}
+ \delta_{\bQ,\bQ_b}\,\frac{\tau_0+(-1)^\ell\,\tau_3}{2},\label{rho_l}
\eal
and thus the Coulomb repulsion can be written as 
\beal
H_C &= \frac{1}{2N\Omega_c}\sum_{\bq,\bG}\,\sum_{\ell\ell'}\,
U_{\ell\ell'}(\bq+\bG)\, \rho^\dagger_\ell(\bq+\bG)\,\rho^\dagga_{\ell'}(\bq+\bG)\,,
\label{H_C}
\eal 
with $N$ the number of supercells, $\Omega_c$ the area of each supercell, $U_{11}(\bq)=U_{22}(\bq)$ and $U_{12}(\bq)=U_{21}(\bq)$
the intra- and inter-layer Fourier transforms of the 
interaction $e^2/r$ screened by the high-frequency dielectric constant 
$\ep_\infty = 9$ of graphene, and by the presence of a dual metal gate~\cite{Zalatel-PRX2020} assumed at distance $30$~nm. \\
The non-retarded attraction mediated by $A_1$ and $B_1$ moir\'e phonons can be straightforwardly derived 
from Ref.~\cite{Mattia-EPJ2020} and is 
\beal
H_P &= -\frac{1}{2\omega_0\,N}\,\sum_{\bq} \,\sum_{a=1,2}\; L_a^\dagger(\bq)\,
L_a^\dagga(\bq)\,,\label{H_p}
\eal
where $\omega_0\simeq 1360~\text{cm}^{-1}$ is the phonon frequency, neglecting its very weak dispersion~\cite{Mattia-JT}, and  
\beal
L_a(\bq) &= \sum_{\sigma\bk}\,\sum_{\bQ\bQ'}\,
\Psi^\dagger_{\bk,\bQ,\sigma}\,\tau_a\,\hat{L}_{\bQ\bQ'}\,\Psi^\dagga_{\bk+\bq,\bQ',\sigma}\,,\label{L_a}
\eal
with
\beal
\hat{L}_{\bQ\bQ'} &= \gamma\,\delta_{\bQ,\bQ'}\,\sigma_0\ + \sum_{i=1}^3\, \big(\delta_{\bQ-\bQ',q_i}+\delta_{\bQ'-\bQ,q_i}\big)\,
\hat{T}_i(g_0,g_1)\,.
\eal
$\hat{T}_i(g_0,g_1)$ are the same as in Eq.~\eqn{T} with $u_0$ and $u_1$ replaced by $g_0$ and $g_1$. We mention that, since sublattices in valleys +1 and -1 are interchanged, $g_0$ and $g_1$ are the modulations induced by the phonons on the intralayer hopping between opposite and equal sublattices, respectively, while 
$\gamma$ refers to the interlayer opposite sublattice one.
The results of realistic tight-binding calculations with frozen phonon displacement~\cite{Mattia-JT} are reproduced by the continuum 
model~\cite{Mattia-EPJ2020} fixing $g_1\simeq g_0/10$ and 
$\gamma\simeq g_0/2.5$, allowing us to parametrise the strength of the phonon-mediated attraction through the single coupling constant   
$g\equiv g_0^2/\omega_0$, with realistic value $\lesssim 1~\text{meV}$.
\\
We remark that $H_0 + H_C$ is invariant under global charge $U(1)$, valley $U_v(1)$ and separate 
spin $SU(2)$ rotations in each valley, thus a large $U(2)\times U(2)$ symmetry~\cite{Zalatel-PRX2020}. On the contrary, the full Hamiltonian $H_0 + H_C + H_P$ is only invariant under $U(1)\times U_v(1)$ 
times the global spin $SU(2)$. The following analysis takes into account just the latter reduced symmetry.

\section*{Results}
\subsection*{Mean-field approximation} 
We perform an all-band Hartree-Fock calculation, thus embracing the full complexity of the band structure and the effects of remote bands on symmetry breaking states. 
Moreover, we take into account the full momentum dependence 
of the Fock term, which ensures robust numerical results. \\
We start analysing the interplay between Coulomb interaction \eqn{H_C} and phonon-mediated attraction \eqn{H_p} at charge neutrality, $\nu=0$, where there is consensus~\cite{Zalatel-PRX2020,Zhang-PRB2020,IV:PhysRevB.103.205414,Liao-PRX2021} that the Coulomb interaction alone stabilises an insulator that has been denoted as Kramer inter-valley coherent (K-IVC) state~\cite{Zalatel-PRX2020}. This is characterised by the order parameter 
\beal
\Delta_\text{K-IVC}(\varphi) \sim \sigma_3\,\big(\cos\varphi\,\tau_1+\sin\varphi\,\tau_2\big)\,,\label{K-IVC}
\eal
which breaks time-reversal symmetry, $\mathcal{T}\sim \tau_1\sigma_1\mathcal{K}$ with $\mathcal{K}$ the complex conjugation, 
and valley $U_v(1)$ symmetry, with generator $\tau_3$, but is invariant under 
$\mathcal{T}\tau_3$. Moreover, it breaks the 
$\mathcal{C}_{2x}\sim \sigma_1$ twofold rotation, while is invariant under the generalised 
\beal
\mathcal{C}'_{2z}(\varphi) \equiv \esp{-i \varphi\tau_3}\;\mathcal{C}_{2z}
\sim \esp{-i \varphi\tau_3/2}\;\tau_1\, \esp{i \varphi\tau_3/2}\,.
\label{C_2z}
\eal
We note that the order parameter \eqn{K-IVC} commutes with the Chern number per spin $\sigma_3$~\cite{Zalatel-PRX2020}. 
Two electrons with given $\sigma_3$ may form a spin-triplet valley-singlet, or a spin-singlet valley-triplet. Eq.~\eqn{K-IVC} implies that Coulomb interaction favours the latter, with valley polarisation $\bd{\tau}$ in the $xy$-plane, opposite for the two different Chern numbers~\cite{Zalatel-PRX2020}.
\\
The phonon-mediated attraction $H_P$ in Eq.~\eqn{H_p}, which can be roughly written as $-g\big(\bd{\tau}\cdot\bd{\tau} -\tau_3^2\big)$, still favours a 
spin-singlet valley-triplet state. However, among the three $\tau_3=-1,0,+1$ components, it lowers the energy of the valley-triplet with $\tau_3=0$ for both $\sigma_3=\pm 1$, thus not breaking any of the symmetries.
This corresponds to a pseudo-rotation in the $\tau_1-\tau_2$ plane: the $U_v(1)$ symmetry is instantaneously broken along a direction in that plane but, on average, dynamically restored. Since these pseudo-rotations describe alternating distortions either along $A_1$ or $B_1$ modes, the state can be regarded as a resonating Kekul\'e valence bond. 
As such, it cannot be represented by a single Slater determinant, hence it is not accessible through the Hartree-Fock (HF). 
%approximation unless we explicitly break $U_v(1)$, 
We can however study by HF the static counterpart once we explicitly break $U_v(1)$ through a static distortion along any  arbitrary combination of the $A_1$ or $B_1$ modes. That amounts to searching for variational wavefunctions with a static Kekul\'e distortion characterised by the two-component order parameter
\beal
\Delta_\text{S-KVB}(\varphi) \sim \sigma_0\,\big(\cos\varphi\,\tau_1  + \sin\varphi\,\tau_2\big)\,,\label{st-JT}
\eal   
which breaks $U_v(1)$, while it is invariant under $\mathcal{C}_{2x}$, $\mathcal{T}$, and the twofold rotation \eqn{C_2z}. 
\begin{figure*}
\centerline{\includegraphics[width=\textwidth]{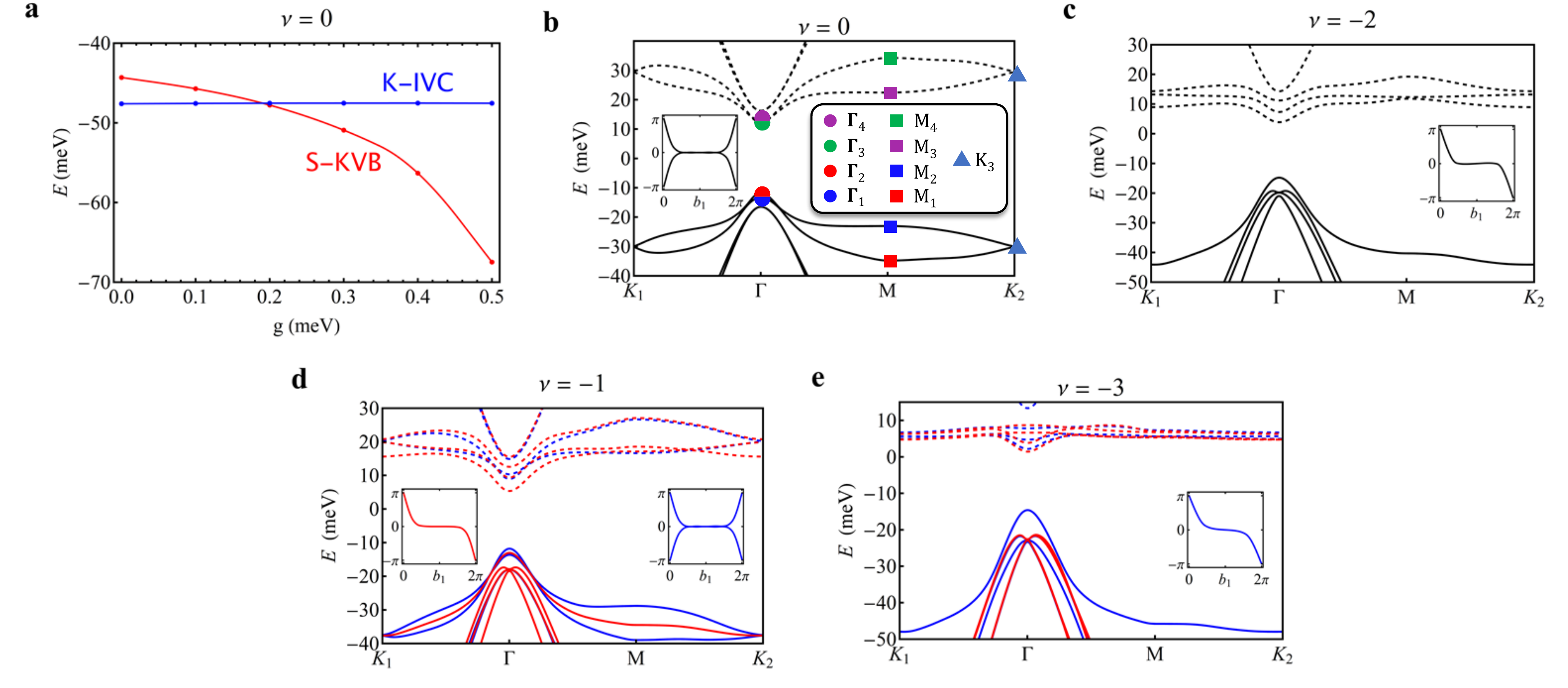}}
\caption{\textbf{Hartree-Fock results}. Panel \textbf{a}: Hartree-Fock energies of the K-IVC and S-KVB variational wavefunctions at charge neutrality, $\nu=0$, characterised, respectively, by  
the order parameters Eq.~\eqn{K-IVC} and Eq.~\eqn{st-JT},
as function of the coupling constant $g$ of the phonon-mediated attraction. 
Panel \textbf{b}: Hartree-Fock band structure of the S-KVB state at $g=0.3~\text{meV}$ and $\nu=0$. Solid lines represent valence bands whereas dashed lines conduction ones.  Also shown are the symmetry properties of the Bloch waves at the high-symmetry points. In the inset, we plot the Wilson loop of the lowest two flat bands.
Panel \textbf{c}: Hartree-Fock band structure of the S-KVB state described by the order parameter Eq.~\eqn{st-JT+T} at $\nu=-2$ and $g=0.3~\text{meV}$. In the inset, we show the Wilson loop of the occupied flat band. 
Panel \textbf{d}-\textbf{e}: Hartree-Fock band structure of the S-KVB state described by the order parameter Eq.~\eqn{st-JT+T}, now spin polarized, at $\nu=-1,-3$ and $g=0.3~\text{meV}$. The blue and red bands correspond to majority and minority spins, respectively. In the inset, the Wilson loops of the occupied bands are shown. The bands for $\nu=1,2,3$ are obtained by a particle-hole transformation.
}
\label{HF-figures}
\end{figure*}
We note that the Fock term of the Coulomb interaction \eqn{H_C} may stabilise either order parameters, although $\Delta_\text{K-IVC}$ is favoured at charge neutrality~\cite{Zalatel-PRX2020}. On the contrary, the phonon mediated attraction \eqn{H_p} only couples via the Hartree term to $\Delta_\text{S-KVB}$. We thus expect that the cooperation between Coulomb and phonon-mediated interactions may eventually make $\Delta_\text{S-KVB}$ prevail over 
$\Delta_\text{K-IVC}$. This is indeed what we find in our calculations. 
In Fig.~\ref{HF-figures}\textbf{a} we show the HF energies of the K-IVC and S-KVB variational states upon increasing the coupling constant $g$ of the attraction. 
At $g=0$, K-IVC is the global minimum and S-KVB a local one.
Increasing $g$, the energy of S-KVB lowers and eventually crosses that of K-IVC. For realistic values of $g$, S-KVB is the stable state, while K-IVC is only metastable. We also took into consideration valley-polarized insulators, which however remain always metastable since they are only weakly coupled to the Fock term of $H_P$, see Section 3 of Supplementary Material~\cite{Supplementary}.
\\
In Fig.~\ref{HF-figures}\textbf{b} we show the HF band structure of the S-KVB state at $g=0.3~\text{meV}$, and indicate explicitly the symmetry properties of the Bloch waves, noting that the space group remains $P622$. The band structure describes an insulator with a sizeable gap $\sim 30~\text{meV}$ separating the two lower flat bands from the upper two. 
The Bloch waves of the lower two bands transform like the irreducible representations $\Gamma_1(1)+\Gamma_2(1)$, $M_1(1)+M_2(1)$ and $K_3(2)$, see Fig.~\ref{HF-figures}\textbf{b}, which hints~\cite{Bernevig-PRL2019} at a `fragile` 
topology, indeed testified by the Wilson loops of the two lower flat bands, see inset of Fig.~\ref{HF-figures}\textbf{b}. In reality, since the S-KVB state is adiabatically connected to the frozen-phonon insulator of Ref.~\cite{Mattia-JT},  
which was shown to support edge states, that topology is actually robust and implies that the two lower flat bands do have finite and opposite Chern numbers $\text{C}=\pm 2$.
\\
For integer fillings away from charge neutrality, the sole $U_v(1)$ symmetry-breaking static Kekul\'e distortion cannot stabilise mean-field insulators, due to the $\mathcal{T}\mathcal{C}_{2z}$ protection of the Dirac cones. Therefore additional symmetries must be broken. 
We can already anticipate how that occurs by noticing that Jahn-Teller coupling is akin inverted Hund's rules~\cite{Capone-PRL2004,Fabrizio-RMP2009,PhysRevB.98.075154} forcing lowest-spin configurations, and that the two occupied flat bands at charge neutrality carry opposite Chern numbers, $\text{C}=\sigma_3=\pm 1$ per spin~\cite{Zalatel-PRX2020}.
Therefore, if phonon contribution prevails over Coulomb exchange, the Chern number degeneracy is split in the first place by a symmetry-breaking term $\propto \sigma_3$, and only as a last resort spin degeneracy is lifted. At even filling $\nu=\pm 2$ this corresponds to spin rotationally invariant topological insulators with spin-singlet order parameter
\be \label{st-JT+T}
\Delta_{\nu=\pm2}(\varphi) \sim \Delta_{\text{S-KVB}}(\varphi) + \sigma_3 \, ,
\ee
breaking time-reversal, $\mathcal{C}_{2x}$ and $\mathcal{C}_{2y}$ symmetries, thus leaving just a $P6$ space group. 
Such topological state is indeed stabilised in mean-field, see Section 3 of Supplementary Material ~\cite{Supplementary}, and its Hartree-Fock band structure at $\nu=-2$ is shown in Fig.~\ref{HF-figures}\textbf{c}. As expected, the occupied flat-band has a nonzero winding number of the Wilson loop, suggestive of a topological Chern insulator with Chern number $\text{C}=\pm2$, consistent with emerging experimental evidences of anomalous quantum Hall effect at $\nu=\pm 2$~\cite{Efetov-arXiv2021,Yankowitz-arXiv2022}. \\
At odd fillings $\nu = \pm 1,\pm 3$, forcing translational symmetry implies that spin degeneracy is unavoidably broken 
by splitting each band with given Chern number $\text{C}$ into two spin-polarised ones with Chern number $\text{C}/2$, which Hartree-Fock indeed does,  see Section 3 of Supplementary Material~\cite{Supplementary}. The band structures 
at $\nu=-1,-3$ are shown in Fig.~\ref{HF-figures}\textbf{d}-\textbf{e}, along with the Wilson loop of the occupied flat bands pointing to a non-trivial topology with $\text{C}=\pm1$.\\
We emphasise that the above results, not in disagreement with experimental evidences~\cite{Sharpe-Science2019,Efetov-SC-Nature2019,Nuckolls2020,Das-NP2021,Pierce-NP2021,PhysRevLett.127.197701,Young-Science2021,Choi-Nature2021,Young-NatPhys2021,Andrei-NatMat2021,Efetov-arXiv2021,Yankowitz-arXiv2022}, depend on Kekul\'e coupling overruling Coulomb exchange~\cite{PhysRevB.98.075154}. This occurs at all integer fillings for $g\sim 0.3 ~\text{meV}$, see Section 3 of Supplementary Material~\cite{Supplementary}.
In the opposite case, K-IVC state would be stable at charge neutrality, and Coulomb exchange should presumably realise conventional Hund's rules, and, therefore, at first lift spin-degeneracy to make high-spin states. That would, e.g., lead to spin $S=1$ non-topological insulators at $\nu=\pm 2$~\cite{Zhang-PRB2020,Zalatel-PRX2020,IV:PhysRevB.103.205414,Supplementary}, contrary to the $S=0$ topological ones that we find; two rather distinct scenarios that can be discriminated experimentally, as well as the two different insulators predicted at charge neutrality.  We mention, for completeness, that there is actually a third possibility we have not taken into account that the insulators at integer $\nu\not=0$ break moir\'e translational symmetry~\cite{Mattia-JT}, which might be stabilised under large enough strain~\cite{Zalatel-PRX2021}.
\\ 
We finally remark that the above results survive a weak $C_{2z}$ symmetry breaking potential in TBLG misaligned to hBN, see Section 4 of Supplementary Material~\cite{Supplementary}, but do not in the case of almost perfect alignment~\cite{Guinea-CM2022}, where clear-cut experimental evidences of insulating states exist only at $\nu=0$ and $\nu=3$~\cite{Serlin-Science2020,Balents-SC-NatPhys2020,Sun2021}.
 
\subsection*{Resonating valence bonds beyond mean field}
Hartree-Fock is only able to describe static Kekul\'e distortions, and predicts tiny atomic displacement because of the large phonon frequency compared to the narrow insulating gaps that are opened.
Since the $A_1$ and $B_1$ moir\'e optical phonon dispersions in momentum space are negligible, around four orders of magnitude less than the center-of-mass frequency~\cite{Mattia-JT}, one can legitimately regard those modes as collective vibrations of a single moir\'e supercell~\cite{Gadelha-Nature2021} as it were a thousand-atom large molecule. 
Therefore, also in light of the extremely narrow width of the flat bands, we cannot exclude that, in reality, Kekul\'e valence bonds resonate, namely they occurs without spontaneously breaking $U_v(1)$.
As we earlier mentioned, that corresponds to the S-KVB distortion, see Fig.~\ref{Kekule}, 
being replaced by a R-KVB one. If that were the case, the above mean-field insulating phases should be replaced by their dynamical counterparts, i.e., by Jahn-Teller Mott insulators~\cite{Jahn-Teller-Mott} in which the effectively inverted Hund's rules and the Coulomb repulsion conspire to halt electron motion and to freeze each moir\'e supercell in the state that maximises the local energy gain with a number of electrons equal to the average one. In the present case of magic-angle TBLG, a simple description of a Jahn-Teller Mott insulator runs into several obstacles. First, each 
supercell contains an unmanageable large number of $\pi$-orbitals that prevents dealing with Jahn-Teller effect as 
one would do in a simple molecule. For that reason, we assume that focusing just on the flat-bands already yields  
a reasonable physical description, in that akin to dealing just with LUMO and HOMO in a molecule. 
That raises another issue: the topological obstruction~\cite{Vishwanath-PRX2018,Po-PRL2018,Bernevig-PRL2019}
prevents building localised Wannier orbitals for the flat bands. To overcome such obstacle, we note that the Jahn-Teller Mott insulator has built-in pairing correlations~\cite{PWA-RVB,Capone-PRL2004,Fabrizio-RMP2009}. With this in mind, we argue that a reasonable description of that state can be gained through a Gutzwiller projected BCS wavefunction~\cite{PWA-RVB}
\beal
\ket{\nu} = P_\text{G}(\nu)\ket{BCS}\,,\label{projected BCS}
\eal
where $\ket{BCS}$ is the BCS wavefunction for the flat bands, and $P_\text{G}(\nu)$ the Gutzwiller projector 
onto the configurations where each supercell is strictly occupied by $4+\nu$ flat-band electrons. Since our goal is just to infer physical features of the resonating counterparts of the S-KVB mean-field insulating states, we shall not attempt to optimise the ansatz wavefunction \eqn{projected BCS}, which is anyhow practically impossible, but assume that its properties are simply inherited by the BCS wavefunction~\cite{Zhang-PRB2013}, hence by the geminal pair-wavefunctions that are favoured by the phonon-mediated attraction and Coulomb repulsion.  
For that, we first project $H_P$ \eqn{H_p} onto the eigenoperators $\Psi^{\dagga}_{\bk,\sigma,\eta,n}$ and $\Psi^{\dagger}_{\bk,\sigma,\eta,n}$ of the flat bands, $n=1$ the lower and $n=2$ the upper, for each valley $\eta=\pm 1$, as previously done for the Coulomb interaction~\cite{Vafek-PRL2019,Zalatel-PRX2020,IV:PhysRevB.103.205414}. Next, we diagonalise the scattering amplitude in the zero-momentum, $\tau_3=0$ and spin-singlet Cooper channel, which is the most favourable one, see Section 5 of Supplementary Material~\cite{Supplementary}, and is 
spanned by the geminal operators 
\be
\Delta^\dagger_{\bk,nm} = \Big(\Pc_{\bk,\up,+1,n} \Pc_{-\bk,\down,-1,m} +
\Pc_{-\bk,\up,-1,m}\Pc_{\bk,\down,+1,n}\Big)/\sqrt{2} \,.\label{geminal}
\ee
The diagonal basis corresponds to the pair creation operators 
\beal
\Delta_{i}^\dagger = \sum_{\bk}\sum_{n,m=1}^2 \, \psi^{nm}_{i}(\bk) \, \Delta^\dagger_{\bk,nm} \, , 
\eal
with eigenvalues $-g\,\lambda_i$ and normalised eigenvectors $\psi^{nm}_{i}(\bk)$. 
\begin{figure}[t]
%\vspace{-3cm}
\centerline{\includegraphics[width=\columnwidth]{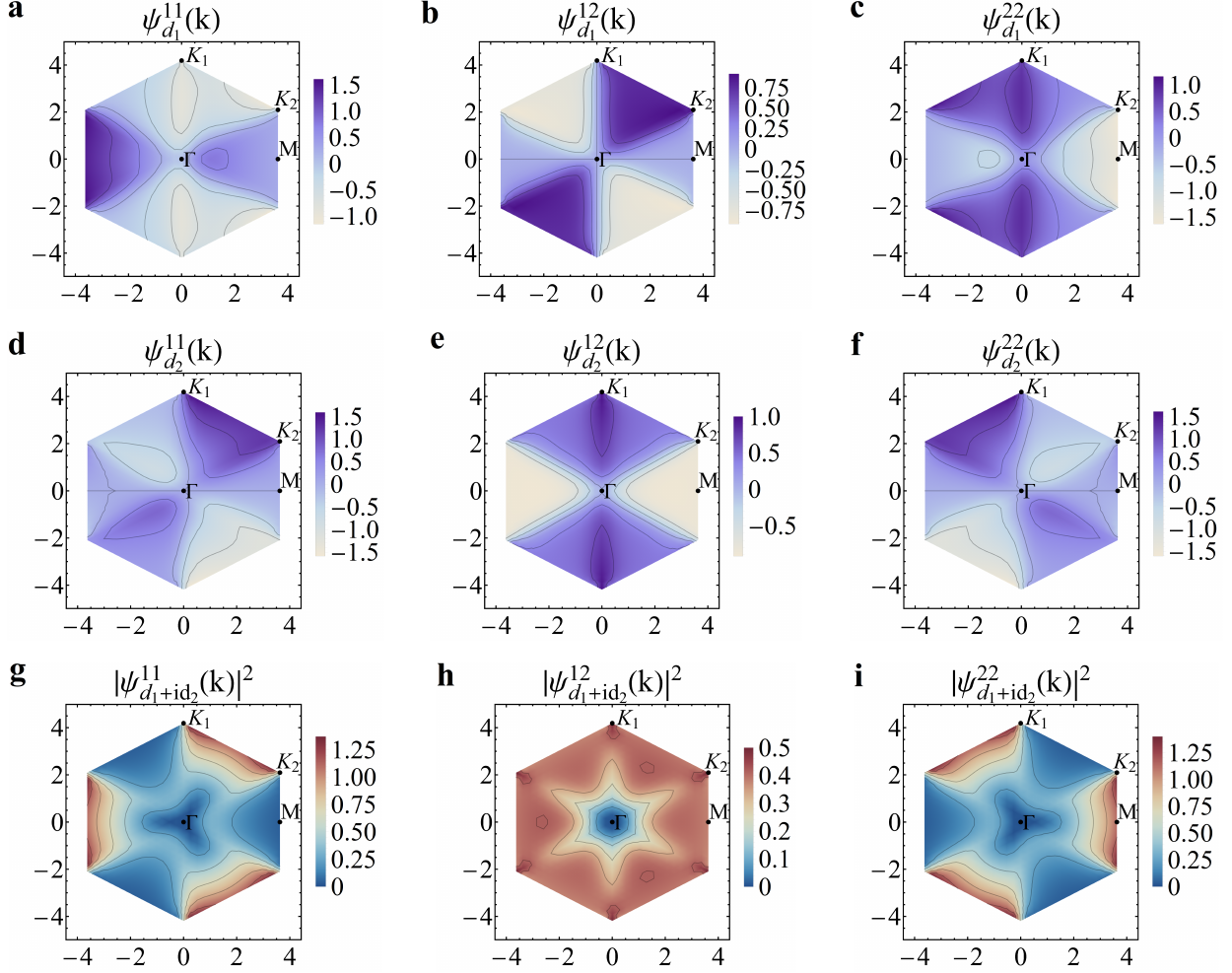}}
%\vspace{-3.5cm}
\caption{\textbf{$d$-wave pair wavefunctions in momentum space}. Panels \textbf{a}-\textbf{f}: pair wavefunction components in momentum space 
of the two $d$-wave real eigenstates $d_1$ and $d_2$, 
namely, $\psi^{11}_{d_1}(\bk)$ (\textbf{a}), $\psi^{12}_{d_1}(\bk)=\psi^{21}_{d_1}(\bk)$ (\textbf{b}), $\psi^{22}_{d_1}(\bk)$ (\textbf{c}), $\psi^{11}_{d_2}(\bk)$ (\textbf{d}), $\psi^{12}_{d_2}(\bk)=\psi^{21}_{d_2}(\bk)$ (\textbf{e}), and $\psi^{22}_{d_2}(\bk)$ (\textbf{f}). These wavefunction are almost odd under $\mathcal{P}$, which implies $\psi^{11}_i(\bk)\simeq -\psi^{22}_i(-\bk)$ and $\psi^{12}_i(\bk)\simeq \psi^{12}_i(-\bk)$.
Panels \textbf{g}-\textbf{h}: probability distribution of the complex $d_1+id_2$ combination, eigenstate of $C_{3z}$ with eigenvalue $\esp{-i2\pi/3}$. The orthogonal combination $d_1-id_2$, eigenstate of $C_{3z}$ with complex conjugate eigenvalue, has right the same probability distribution.}
\label{d-wave}
\end{figure}
The transformation rules of $\psi_i^{nm}(\bk)$ under the symmetry transformations 
are thoroughly discussed in the Section 5 of Supplementary Material~\cite{Supplementary}. Here we just remark that 
the component of 
$\psi^{nm}_i(\bk)$ even under inversion $\bk\to-\bk$ corresponds to the spin-singlet, valley-triplet with $\tau_3=0$, 
while the component odd under inversion  to the spin- and valley-singlet. Both components have $\tau_3=0$ and therefore their mutual coupling is allowed by $U_v(1)$.
\\
By construction, the lowest-energy pair-eigenstate is the one with the largest $\lambda_i$.
We find that the largest eigenvalue $\lambda_s=2.33$ is non-degenerate and its eigenvector, i.e., the pair wavefunction, 
transforms like the totally symmetric irreducible representation of $D_6$, which is alike an $s$-wave Cooper pair. 
The next largest eigenvalue $\lambda_{d}=1.79$ is doubly degenerate, and the corresponding pair wavefunctions transform like the two-dimensional irreducible representation of $D_6$ even under $C_{2z}$, alike a $d$-wave Cooper pair. \\
However, we have so far just considered the phonon-mediated attraction that, unsurprisingly, favours the $s$-wave pairing channel. That result may change taking into account also the Coulomb repulsion $H_C$ in Eq.~\eqn{H_C}. However, a proper treatment of $H_C$ would require including all thousands $\pi$-bands, because the Coulomb repulsion projected just onto the flat bands, i.e.,  without the screening by all other bands, is unphysical. Since that calculation is not feasible, we just compute the difference $\Delta \mu_*$ of the Coulomb pseudo-potentials between $d$- and $s$-wave pairs, with the Coulomb repulsion screened within RPA by all $\pi$-bands but the flat ones, see Section 7 of Supplementary Material~\cite{Supplementary}. We find $\Delta \mu_* \simeq -0.49~\text{meV}$, which implies that the $d$-wave channel is the lowest energy one for electron-phonon coupling $g \lesssim 0.88~\text{meV}$ in presence of the Coulomb repulsion~\cite{MacDonald-phonons-PRL2018}.
\\
Since the interplay of the phonon-mediated attraction and Coulomb repulsion stabilizes the $d$-wave pairing, we hereafter just focus on the latter. In the real representation, one eigenstate $\psi^{nm}_{d_1}(\bk) \sim x^2-y^2$, even under $C_{2x}$, and the other $\psi^{nm}_{d_2}(\bk)\sim xy$, odd under $C_{2x}$, see Fig.~\ref{d-wave}. In reality, since $\psi^{nn}_{d_i}(\bk)\not=\psi^{nn}_{d_i}(-\bk)$, each eigenstate has also a weak $p$-wave valley-singlet component, $p_x$ and $-p_y$ the eigenstates $d_1$ and $d_2$, respectively.
Both $d_1$ and $d_2$ are (almost) odd under p-h symmetry, as can be noticed in 
Fig.~\ref{d-wave}, which hints at a non-trivial topological character~\cite{II:PhysRevB.103.205412,III:PhysRevB.103.205413}. Indeed, the combinations 
$d_\pm = \big(d_1 \pm i\,d_2\big)/\sqrt{2}\sim Y_{2\pm 2}$ do have finite Chern number $\text{C}=\pm 2$, the same value of the angular momentum, see Section 6 of Supplementary Material~\cite{Supplementary}.
\begin{figure}
%\vspace{-3cm}
\centerline{\includegraphics[width=0.8\columnwidth]{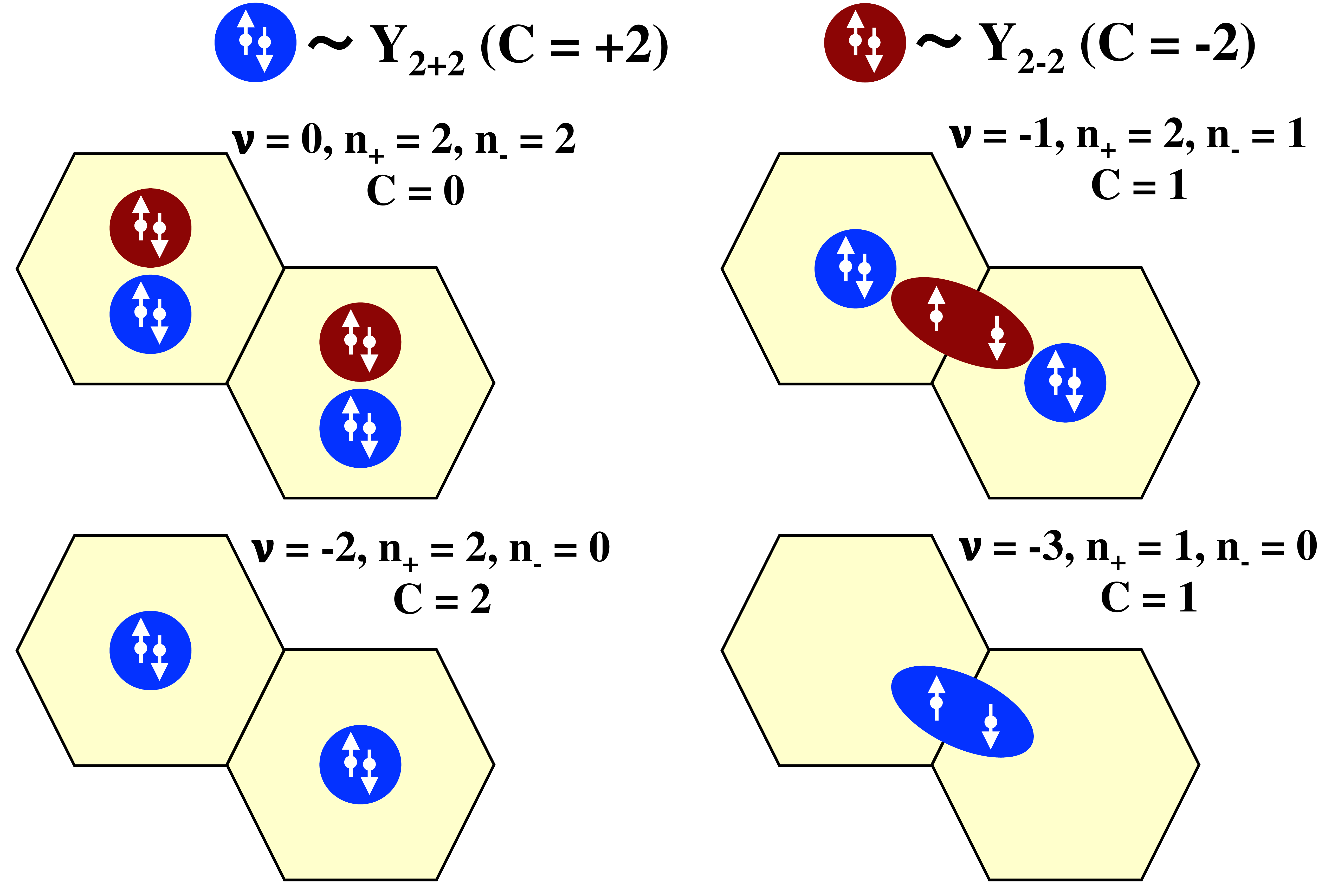}}
%\vspace{-3.5cm}
\caption{\textbf{R-KVB insulators at integer $\mathbf{\nu\leq 0}$}. The two spin-singlet, $\tau_3=0$ geminal operators: 
$\Delta^\dagger_+\sim Y_{2+2}$ 
with Chern number $C=+2$ is represented by the blue circle, while $\Delta^\dagger_-\sim Y_{2-2}$ 
with $C=-2$ by the red one. A pictorial representation of 
the resonating Kekul\'e valence bond insulators, dynamical counterparts of the static mean-field 
ones in Figs.~\ref{HF-figures}, is displayed for all negative integer fillings. For each case, we draw two nearest neighbour moir\`e unit cells, each of which, for odd $\nu$, hosts one electron from the shared pair; a very oversimplified picture of spin- and valley-liquid insulators.}
\label{summary}
\end{figure}
\noindent

Therefore, under the above assumptions, and, at first instance, neglecting the flat band dispersion, R-KVB insulators can be stabilised at all integer fillings $\nu$. These are described by the projected BCS wavefunctions, see Eq.~\eqn{projected BCS}, 
\beal
\ket{\nu} &\propto P_\text{G}(\nu)\,\Big(\Delta_{d_+}^\dagger\Big)^{\frac{N}{2} \, n_+}
\Big(\Delta_{d_-}^\dagger\Big)^{\frac{N}{2} \, n_-}\ket{0}\,,
\label{any nu}
\eal
where, by definition, the `vacuum' $\ket{0}$ is the ground state state at $\nu=-4$ with all bands below the flat ones occupied, and $n_\pm \geq 0$ are the numbers of $d_\pm$ pairs per unit cell. The filling factor is simply $\nu=-4+n_+ + n_-$. 
Whenever $n_+\not=n_-$, the projected wavefunction \eqn{any nu} breaks time-reversal symmetry, carries Chern number $C(n_+,n_-) = (n_+-n_-)$, and has finite orbital magnetic moment $M=\mu_B\,g_*\,C(n_+,n_-)$ per supercell, with $g_*$ the gyromagnetic ratio. 
Since the explicit calculation~\cite{Resta} of the gyromagnetic ratio is unfeasible in our scheme, we rely on the experimental estimate of $g_*\sim 3$~\cite{Efetov-arXiv2021}, yielding an orbital magnetic moment per pair of $\sim 6$ Bohr magnetons.  
\\ 
The dynamical counterparts of the S-KVB mean-field 
insulators Figs.~\ref{HF-figures}\textbf{b}-\textbf{e} correspond to specific values of $n_+$ and $n_-$, see Fig.~\ref{summary}.
However, due to the large value of $g_*$ it is well possible that pairs $(n_+,n_-)$ with Chern number higher in absolute value than the mean-field solutions could become stable in presence of a magnetic field, possibly realising the peculiar Landau fan diagrams that have been observed~\cite{Nuckolls2020,Pierce-NP2021,PhysRevLett.127.197701,Das-NP2021,Young-NatPhys2021,Choi-Nature2021,Andrei-NatMat2021}. \\
We note that, since at odd integer $\nu$ the R-KVB cannot quench spin and valley degrees of freedom, the wavefunction \eqn{any nu} describes in that case spin- and valley-liquid topological insulators, whereas Hartree-Fock predicts fully-polarised symmetry breaking ones. 
Nonetheless, that wavefunction still has finite orbital magnetisation, which, e.g., could be as large as $3\mu_B$ per moir\`e supercell at $\nu=3$, not in disagreement 
with recent observations~\cite{Young-Science2021}. Moreover, the sizeable 
orbital magnetic moment implies the emergence of magnetic domains at any integer filling $\nu\not=0$ rather than a uniform magnetic polarisation. Since 
the orbital magnetic moment of each pair is directly proportional to its Chern number, that 
envisages the existence of domains with different Chern numbers, as indeed observed 
experimentally~\cite{Efetov-mosaic2022}. 
We further remark that R-KVB insulators are prone to turn upon doping  into 
superconductors~\cite{Capone-PRL2004,Fabrizio-RMP2009}, in the present case
nodeless chiral $d$-wave ones that are still topological~\cite{Liu-PRL2018}, whose driving mechanism, we emphasise, 
is the electron-phonon Kekul\'e coupling~\cite{science.abb8754}.\\
Since HF predicts time-reversal symmetry-breaking topological insulator at $\nu \neq 0$, that suggests that the complex pair operators $\Delta^\dagger_{\pm}$ prevail over their real combinations. That is presumably consequence of Coulomb exchange, though we cannot exclude to be just a mean-field artefact. Therefore, for completeness, let us briefly discuss what would change if  
$\Delta^\dagger_{1}$ and $\Delta^\dagger_{2}$ were instead favoured. 
In that case, we have simply to replace 
$\big(\Delta_\pm^\dagger,n_\pm\big)$  
with $\big(\Delta_{1(2)}^\dagger,n_{1(2)}\big)$ in 
the wavefunction \eqn{any nu}, which would thus describe non-topological R-KVB insulators with a weak nematic character due to the small $p$-wave component. Moreover, the symmetry of the superconducting order parameter stabilised upon doping would be now a real combination of $d_{x^2-y^2}$, plus a small $p_x$ component, and $d_{xy}$, plus a small $-p_y$ component, implying nodes in the Brillouin zone~\cite{Tan-NAT2021} and weak nematicity~\cite{p+d-NPJ2019}. 

\section*{Conclusions} 
The surprisingly rich phase diagram of magic-angle twisted bilayer graphene, which includes topological and non-topological correlated 
insulators~\cite{Herrero-MIT-Nature2018,Sharpe-Science2019,Efetov-SC-Nature2019,Nuckolls2020,Das-NP2021,Pierce-NP2021,Young-Science2021,Choi-Nature2021,Young-NatPhys2021,Andrei-NatMat2021}, sometimes competing with each other~\cite{PhysRevLett.127.197701}, and superconducting domes~\cite{Herrero-SC-Nature2018,Efetov-SC-Nature2019,Efetov-arXiv2021,Yankowitz-arXiv2022}, is explained by the constructive interplay of the Coulomb repulsion and the effective attraction mediated by a rather peculiar set of moir\'e optical phonons.
The Kekul\'e-like valence-bond state, only metastable in presence of just the Coulomb repulsion, is stabilized by this interplay 
and characterized by a distortion localised mostly into the AA regions and along the domain walls separating AB and BA Bernal Stacked regions of twisted bilayer graphene. The presence of such particular Kekul\'e distortion could be determined by a combination of high resolution STM and Chern number measurements \cite{PhysRevLett.129.117602}.
The resulting physical scenario is in agreement with the experimentally observed insulating states at all integer fillings. Moreover, it naturally offers an explanation of the observed superconductivity~\cite{science.abb8754} and its proximity to the insulating phases~\cite{Capone-PRL2004,Fabrizio-RMP2009}.

\section*{Acknowledgments}
We are grateful to Mattia Angeli and Erio Tosatti for helpful discussions and comments.
We acknowledge funding from the European Research Council (ERC), under the European Union's Horizon 2020 research and innovation programme, Grant agreement No. 692670 ``FIRSTORM'', and from Italian Ministry of University and Research under the PRIN 2020 programme, project No. 2020JLZ52N.

\bibliographystyle{apsrev4-2}
\bibliography{mybiblio}

\end{document}